\documentclass[1p,preprint,times]{elsarticle}
\usepackage{amssymb,amsmath,placeins,epsfig,array}
\usepackage[usenames,dvipsnames]{color}
\usepackage[normalem]{ulem}
\usepackage[breaklinks,colorlinks,urlcolor=blue,citecolor=blue,linkcolor=blue]{hyperref}
\usepackage{float}
\usepackage{mathptmx}
\usepackage{physics}
\usepackage{bm}
\journal{NIM A}

\usepackage{scalerel}
\usepackage{tikz}
\usetikzlibrary{svg.path}

\definecolor{orcidlogocol}{HTML}{A6CE39}
\tikzset{
	orcidlogo/.pic={
		\fill[orcidlogocol] svg{M256,128c0,70.7-57.3,128-128,128C57.3,256,0,198.7,0,128C0,57.3,57.3,0,128,0C198.7,0,256,57.3,256,128z};
		\fill[white] svg{M86.3,186.2H70.9V79.1h15.4v48.4V186.2z}
		svg{M108.9,79.1h41.6c39.6,0,57,28.3,57,53.6c0,27.5-21.5,53.6-56.8,53.6h-41.8V79.1z M124.3,172.4h24.5c34.9,0,42.9-26.5,42.9-39.7c0-21.5-13.7-39.7-43.7-39.7h-23.7V172.4z}
		svg{M88.7,56.8c0,5.5-4.5,10.1-10.1,10.1c-5.6,0-10.1-4.6-10.1-10.1c0-5.6,4.5-10.1,10.1-10.1C84.2,46.7,88.7,51.3,88.7,56.8z};
	}
}

\newcommand\orcidicon[1]{\href{https://orcid.org/#1}{\mbox{\scalerel*{
				\begin{tikzpicture}[yscale=-1,transform shape]
					\pic{orcidlogo};
				\end{tikzpicture}
			}{|}}}}
		
\begin{document}
	
	\begin{frontmatter}
		
		\title{Simulation of the spatial shift in detector response for polarized protons within a calorimeter}
		
		\author[UNC]{A.~Blitstein\corref{cor}}
		\cortext[cor]{Tel.: +1-704-792-8801}
		\ead{austinmb@live.unc.edu}
		\author[JLAB]{B.~Wojtsekhowski}
		
		\address[UNC]{\mbox{Department of Physics and Astronomy, University of North Carolina, Chapel Hill, North Carolina, 27599, USA}}
		\address[JLAB]{\mbox{Thomas Jefferson National Accelerator Facility{,} Newport News{,} Virginia 23606, USA}}
		
		\begin{abstract}
			Measurement of the helicity dependent elastic electron-proton scattering cross section provides a key means of investigating parity violation within the proton. However, such measurements exhibit potential instrumental effects associated with the detection of polarized recoiled protons. In particular, spin-orbit interactions within a massive detector induce a systematic spatial shift in the detector signal. 
			In this study, we determine the size of this shift using the Geant4 simulation toolkit. For a typical hadron calorimeter, we found a polarization dependent shift on the order of 0.01-0.1 mm, multiple orders of magnitude smaller than the typical spatial resolution seen in hadronic calorimeters. 
			Additionally, we provide the custom modifications required of the Geant4 source code to implement the quasi-elastic scattering of polarized protons incident on nuclei in the detector. 
			The modifications are readily extendable to generic matter sources, and can be used for the study of additional spin dependent observables in Geant4.
		\end{abstract}
		
	\end{frontmatter}
	
	\section{Introduction}
	
	Much of what is currently known about the structure of the proton has been ascertained from electron scattering experiments~\cite{Hofstadter1956,Friedman1972,Ji2021,Punjabi2015}. 
	In such experiments, a beam of electrons is incident on a proton target. 
	By measuring the distribution of kinematic variables associated with the scattered particles, such as their momentum and energy, one obtains data on various proton form factors. 
	These form factors are then used to infer the spatial distribution of quarks and gluons within the proton.
	In this way, the use of polarized beams and polarized targets allow one to probe the origin of the spin of the nucleon.
	
	The observation of the small value of the quark polarization in the nucleon (``spin crisis")~\cite{ADAMS1994399} raises a question about the strange quark contribution to the spin of the nucleon.
	The parity violation present in elastic electron-proton scattering allows for the determination of the proton form factors related to the strange quark~\cite{Beck-McKeown2001}. 
	In such experiments the cross section beam helicity asymmetry is measured. 
	The detected particles are also polarized, which could require systematic corrections to the measurement results on account of spin-dependent interactions within the detector. 
	These instrumental effects become significant for the detection of polarized recoiled protons. 
	
	Consequently, it is necessary to ensure that these spin dependent interactions that occur within the detector do not collude to disrupt the determination of the coordinate measurement of the polarized protons by the detector. It is known that polarized protons are affected by a spin-orbit interaction when scattering off nuclei in matter~\cite{walter_analyzing_1997,jackson}. 
	This spin-dependent, quasi-elastic scattering can be accounted for via the introduction of an empirically known asymmetry in the azimuthal distribution of scattered protons. 
	In many cases, this data is already available. 
	Here, we make use of data~\cite{azhgirey_measurement_2005} for hydrocarbon based targets obtained with a standard hadron sampling calorimeter.
	
	The specific motivation for the current study is a possible experiment proposed in Ref.~\cite{wojtsekhowski_flavor_2020}.
	In what follows, we present a polarization-dependent prediction for the expected size of the spatial shift in detector signal induced by spin-orbit interactions within the proposed calorimeter.
	To do so, simulations were run in Geant4, a Monte-Carlo based toolkit for simulating nuclear physics that is sourced in C++~\cite{agostinelli_geant4simulation_2003}. 
	Currently, generic Geant4 has no default implementation for many of the spin-dependent interactions that are necessary to accurately simulate parity violating effects in nuclear physics. 
	Consequently, it was necessary to introduce custom modifications to the source code of Geant4 associated with the modeling of the elastic scattering of protons incident on nuclei in matter. 
	This implementation is readily generalizable to account for alternative modifications to the azimuthal part of the scattering statistics, and can be used in further Geant4 studies involving spin dependent observables (such as those seen in Ref.~\cite{PERDRISAT2007694}).
	
	\section{The Propagation of Polarized Protons in Matter}
	\label{sec:PPM}
	Due to their interaction with the incident beam of polarized particles, the protons that undergo elastic scattering with said particles are polarized as well. 
	In particular, for elastic $e - p$ scattering in the single photon exchange approximation, the transverse polarization $P_{t}$ of the recoiled protons for a longitudinally polarized electron beam is:
	\begin{equation}
		P_t = -2h\,P_e \sqrt{\tau(1+\tau)} \,\, \tan (\frac{\theta_e}{2}) 
		\frac{G_{_E} \cdot\, G_{_M}}{G_{_E}^2 + (\tau/\epsilon) G_{_M}^2},
		\label{eq:Pt}
	\end{equation}
	where $P_{e}$ is the electron beam polarization, $h = \pm 1$ the beam helicity, $\theta_{e}$ the electron scattering angle, $\tau = Q^{2}/4m_p^{2}$ \hskip .05 in ($Q^2$ the negative four-momentum transfer squared, $m_p$ the proton mass), $\epsilon$ the virtual photon polarization parameter, and $G_{_E}$ and $G_{_M}$ the electric and magnetic form factors~\cite{Punjabi2015}.
	
\begin{figure}[ht!]
		\centering
		\includegraphics[width=\linewidth]{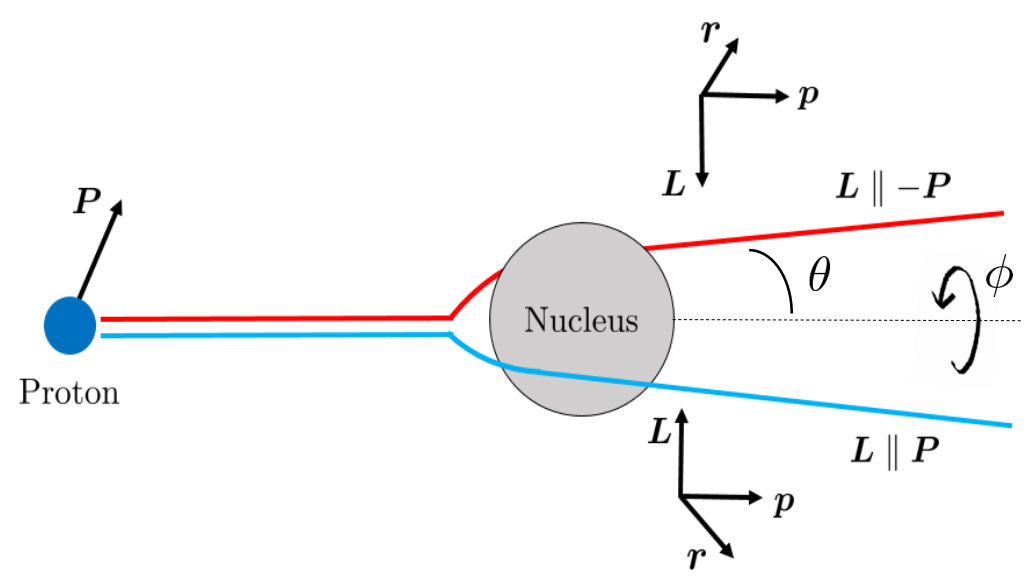}
		\caption{Diagram illustrating the spin-orbit interaction felt by polarized protons elastically scattering off nuclei in matter. It is known empirically that protons prefer to take the path with $\bm{L}\parallel -\bm{P}$, indicated by the top/red trajectory. Color online.}
		\label{fig:spin-orbit}
	\end{figure}

	To first order, when polarized protons elastically scatter off nuclei in matter, they experience a spin-orbit interaction~\cite{walter_analyzing_1997,jackson}. 
	Depending on the path they take around a given nucleus, the direction of their angular momentum relative to the nucleus differs, as seen in Fig.~\ref{fig:spin-orbit}. 
	If the spin orbit interaction is such that it prefers anti-alignment of proton spin with its angular momentum, it will preferentially scatter off the nucleus in the direction that fulfills that requirement (the top/red trajectory in Fig.~\ref{fig:spin-orbit}). 
	For polarized protons in matter, it is empirically known that protons do preferentially scatter in the direction that promotes anti-alignment of their spin and angular momentum, the same direction as their spin axis crossed with their incident direction (see Ref.~\cite{walter_analyzing_1997} for a summary of the spin-orbit interaction for nucleon-nucleus scattering).
	
	The elastic scattering of polarized protons incident on spinless nuclei depicted in Fig.~\ref{fig:spin-orbit} has a few key features which must be considered when modeling the interaction. First, the spin orbit piece to the interaction should not modify the distribution of the scattering angle, $\theta$, for scattered protons. This piece of the scattering statistics reflects the typical treatment of protons elastically scattering off nuclei, providing an important benchmark that our modified interaction should reproduce. 
	Accordingly, all that will be changed is the distribution associated with the azimuthal angle, $\phi$, which sets the scattering direction in the plane normal to the incident direction.
	
	For standard proton-nucleus elastic scattering, the azimuthal distribution is taken to be uniform, with no preference in direction for scattering. To add in such a preference, we will need to introduce an asymmetry in the azimuthal distribution for scattered protons. 
	Taking the z-axis to be along the incident direction and proton spin to be measured along the y-axis, the scattering angular distribution takes the form
	\begin{equation}
		N(\theta,\phi) = N_{0}(\theta)\left(1 + A_{y}P_{y}\cos\phi\right),
		\label{eq:N-azimuthal}
	\end{equation}
	where $N_{0}(\theta)$ is the angular distribution in the absence of proton polarization~\cite{azhgirey_measurement_2005}. 
	The side (transverse) polarization, $P_{y}$, is defined as
	\begin{equation}
		P_{y} = \frac{N_{y}^{\uparrow} - N_{y}^{\downarrow}}{N_{y}^{\uparrow} + N_{y}^{\downarrow}},
	\end{equation}
	where $N_{y}^{\uparrow/\downarrow}$ is the total number of spin-up/down protons with respect to the y-axis. 
	One can also view the side polarization as twice the average y-component of incident proton spin. 
	The other variable that appears in Eq.~\ref{eq:N-azimuthal}, the analyzing power $A_{y}$, is there for the sole purpose of accounting for any additional azimuthal dependence seen in experiments/theory. 
	
	For our purposes, we are primarily interested in the elastic scattering of protons incident on nuclei within a plastic scintillator. 
	For protons incident with lab frame momentum $\vec{p}_{\text{lab}} = p_{\text{lab}} \,\, \hat{z}$, it is known from experiment that the analyzing power is well approximated by
	\begin{equation}
		A_{y}(\theta) = \frac{\sum_{n=1}^{4}c_{n}\left(p_{\text{lab}}\sin\theta\right)^{n}} {p_{\text{lab}}},
		\label{eq:Ay}
	\end{equation}
	where $c_{1} = 3.02 \pm 0.13$, $c_{2} = -7.33 \pm 0.66$, $c_{3} = 6.17 \pm 1.11$, and $c_{4} = -1.74 \pm 0.59$, with $p_{\text{lab}}$ in GeV/c (see Ref.~\cite{azhgirey_measurement_2005}).
	
	Looking at Eq.~\ref{eq:N-azimuthal}, we see that the effect of the added asymmetry in the azimuthal distribution is to preferentially scatter protons in the direction corresponding to $\phi = 0$ (where the extra term has a maximum) while making it less likely to scatter in the direction with $\phi = \pm\pi$ (where the extra term has a minimum). Further, the additional factor averages over $\phi\in(-\pi,\pi)$ to 1, thereby not changing the net scattering angle part of the angular distribution.
	
	\section{Geant4 Custom Modifications}
	
	To model the elastic scattering of polarized protons in matter, simulations were run in Geant4. 
	Geant4 is a toolkit composed of C++ source files intended for simulating nuclear physics via Monte Carlo techniques (see Ref.~\cite{agostinelli_geant4simulation_2003} for more than what is presented here). 
	Interactions between particles are handled in a generic way by associating with each an interaction length that goes as the inverse of the interaction cross section, thereby allowing for modular adjustments to its code. 
	
	A generic Geant4 simulation is called a run, which is composed of a specified number of events, or repetitions of the user defined initial conditions. Each event is then split up into multiple tracks for each of the primary particles and any secondaries that are generated. 
	Then, each track is updated in small increments called steps, implementing dynamic changes to each of the particles with a mix of at rest (e.g. radioactive decay), continuous (e.g. ionization), and discrete (e.g. pair production) processes.
	
	The typical Geant4 user need only modify a few concrete instances of the core base classes which guide the workflow of a typical simulation. Changes to detector geometry and materials are handled in a concrete instance of the \textit{G4VUserDetectorConstruction} class. 
	Particle definitions and their interactions are set in an instance of the \textit{G4VUserPhysicsList} class. 
	Finally, event initialization, such as setting the initial particles and their kinematic properties, is specified in an instance of the \textit{G4VUserActionInitialization} class. 
	Though the aforementioned classes are the only mandatory classes the user should create, additional classes can be defined, which allow one to interface with the particles at the beginning and end of each run, event, track, and step. 
	These are set in concrete instances of the \textit{G4UserRunAction}, \textit{G4UserEventAction}, \textit{G4UserTrackingAction}, and \textit{G4UserSteppingAction} base classes respectively.
	
	While users usually only need to associate predefined physics processes with the particles they involve, this fails when Geant4 does not have a default implementation for the interaction of interest. As it turns out, this is the case for the elastic scattering of polarized protons incident on nuclei. To implement the interaction, it is necessary to directly modify the source code containing the default implementation of hadron elastic scattering, \textit{G4HadronElasticProcess.cc}.
	
	Within the default code, the azimuthal angle is chosen randomly from -$\pi$ to $\pi$. 
	In order to account for the asymmetry present in Eq.~\ref{eq:N-azimuthal}, we instead need to sample $\phi$ from the following normalized probability density function
	\begin{equation}
		F(\phi) = \frac{1}{2\pi}\left(1 + A_{y}P_{y}\cos\phi\right).
		\label{eq:F}
	\end{equation}
	In the modified code, this is accomplished by first integrating $F(\phi)$ to get the cumulative distribution function,
	\begin{equation}
		Q(\phi) = \int_{-\pi}^{\phi}F(\phi')\dd{\phi'} = \frac{1}{2} + \frac{\phi}{2\pi} + A_{y}P_{y}\sin\phi.
		\label{eq:Q}
	\end{equation}
	Then, a random sample $\phi_{s}$ from $F(\phi)$ can be obtained by first taking a random sample $x_{s}$ from the uniform distribution on $[0,1]$, $\mathcal{U}(0,1)$, followed by solving the implicit equation
	\begin{equation}
		x_{s} = Q(\phi_{s}) = \frac{1}{2} + \frac{\phi_{s}}{2\pi} + A_{y}P_{y}\sin\phi_{s},
		\label{eq:x-s}
	\end{equation}
	for $\phi_{s}$ such that $\phi_{s} = Q^{-1}(x_{s})$. 
	In the modified code, Eq.~\ref{eq:x-s} is solved using the bisection method with an accuracy of $\pi / 2^{11}$. We find the sampling of $\phi_{s}$ from $F(\phi)$ via the bisection method induces a 10\% increase in computation time relative to the default code. As we only require a few days of computation time to produce sufficiently small systematic uncertainty, we choose not to employ a more efficient sampler such as the standard rejection method.
	
\begin{figure}[ht!]
		\centering
		\includegraphics[width=\linewidth]{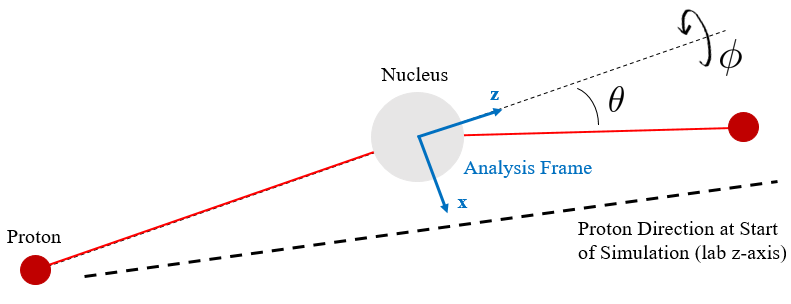}
		\caption{Depiction of the analysis frame used to set scattering angles in Geant4. Color online.}
		\label{fig:2}
	\end{figure}

	Though we now have a procedure to sample $\phi$ from an asymmetrical distribution, we need to be careful about what azimuth $\phi = 0$ corresponds to. In Sec.~\ref{sec:PPM}, this was the azimuth of the cross product of the incident proton's spin axis with its incident direction. Geant4, however, has a different process for assigning this azimuth, one that rotates the coordinate system that the scattering analysis is performed in differently each time. This analysis frame, in which the scattering angles are set, is defined to be the right handed coordinate system with the z-axis along the direction of propagation of the proton immediately before the scattering event, and the x-axis chosen such that the proton's direction at the start of the simulation is contained within the xz plane of this new coordinate system. This still leaves us with two possible choices for the direction of the x-axis, so the unique analysis frame is specified as the one in which the inner product of the unit vector along the x-axis with the proton direction at the start of the simulation is positive. We note that the proton direction at the start of the simulation and the proton direction immediately before the scattering event will generally differ due to continuous Geant4 processes. Refer to Fig.~\ref{fig:2} for an example of this choice of analysis frame.
	
	In Geant4, three vectors are defined as instances of the \textit{G4ThreeVector} class. To rotate vectors from the analysis frame into the default lab frame, the public member function \textit{rotateUz(indir)} is used, where \textit{indir} is a normalized G4ThreeVector pointing in the direction of the proton immediately before elastic scattering. 
	To determine the azimuth corresponding to $\phi = 0$ in the analysis frame, we will need to transform the direction along which the spin of the proton is measured from the lab frame into the analysis frame. This is encoded by the inverse of the rotation accomplished by \textit{rotateUz(indir)}, which had to be added into the code as it was not one of the default G4ThreeVector member functions. More specifically, if $\bm{v} = (v_{1}, v_{2}, v_{3})^{\text{T}}$ is a vector and \textit{indir} = $(u_{1}, u_{2}, u_{3})^{\text{T}}$ a unit vector, we can represent the action of \textit{rotateUz(indir)} on $\bm{v}$, \textit{v.rotateUz(indir)}, as
	\begin{equation}
		\mqty(\frac{u_{1}u_{3}}{u_{1}^{2} + u_{2}^{2}} & -\frac{u_{2}}{u_{1}^{2} + u_{2}^{2}} & u_{1} \\
		\frac{u_{2}u_{3}}{u_{1}^{2} + u_{2}^{2}} & \frac{u_{1}}{u_{1}^{2} + u_{2}^{2}} & u_{2} \\ -u_{1}^{2}-u_{2}^{2} & 0 & u_{3})\mqty(v_{1} \\ v_{2} \\ v_{3}) = \bm{R}\bm{v}
		\label{eq:matr1}
	\end{equation}
	upon which we added in the inverse rotation \textit{rotateUzInv(indir)} as follows:
	\begin{equation}
		\mqty(\frac{u_{1}u_{3}}{u_{1}^{2} + u_{2}^{2}} & \frac{u_{2}u_{3}}{u_{1}^{2} + u_{2}^{2}} 
		& -u_{1}^{2}-u_{2}^{2} \\ -\frac{u_{2}}{u_{1}^{2} + u_{2}^{2}} & \frac{u_{1}}{u_{1}^{2} + u_{2}^{2}} & 0 \\ u_{1} & u_{2} & u_{3})\mqty(v_{1} \\ v_{2} \\ v_{3}) = \bm{R}^{-1}\bm{v}.
		\label{eq:matr2}
	\end{equation}
	
	Since we take proton polarization to be measured along the y-axis of the lab frame, we first get the components of the $\hat{y}$ unit vector in the analysis frame using Eq.~\ref{eq:matr2}. 
	Taking a cross product of this vector with a unit vector along the z-axis in the analysis frame then gives a vector who's azimuth corresponds to $\phi = 0$ when sampling from Eq.~\ref{eq:F}. 
	To take this into account, we find the current azimuthal angle of this new vector in the analysis frame, add it to our sampled $\phi_{s}$ from before, and set that as the new azimuthal angle in the analysis frame. 
	After some (default) post processing, the scattered proton momentum is rotated back into the lab frame through Eq.~\ref{eq:matr1} and changes to the kinematic properties of the particles involved are set by the Geant4 code.
	
	In order to actually carry out these changes, one must ensure they are changing the G4ThreeVector \textit{outdir} that is being passed to the \textit{ProposeMomentumDirection()} member function associated with the variable \textit{theTotalResult}, which gets returned by the \textit{G4HadronElasticProcess::PostStepDoIt()} process. 
	Accordingly, the changes to \textit{outdir} can be made anywhere between its initial definition and setting of the proposed momentum direction in the default \textit{G4HadronElasticProcess.cc} source code.
	
	Besides the modified hadron elastic scattering code, the other standard interactions and particles needed to simulate protons incident on calorimeters are included. In particular, we use the physics lists defined within the default \textit{FTFP\_BERT.cc/.hh} files. 
	Doing so accounts for electromagnetic and synchroton physics, particle decays, hadronic physics, stopping physics, ion physics, and cuts for tracking neutrons.
	
	\section{Benchmarks with Constant Analyzing Power}
	
\begin{figure}[ht!]
		\centering
		\includegraphics[width=0.85\linewidth]{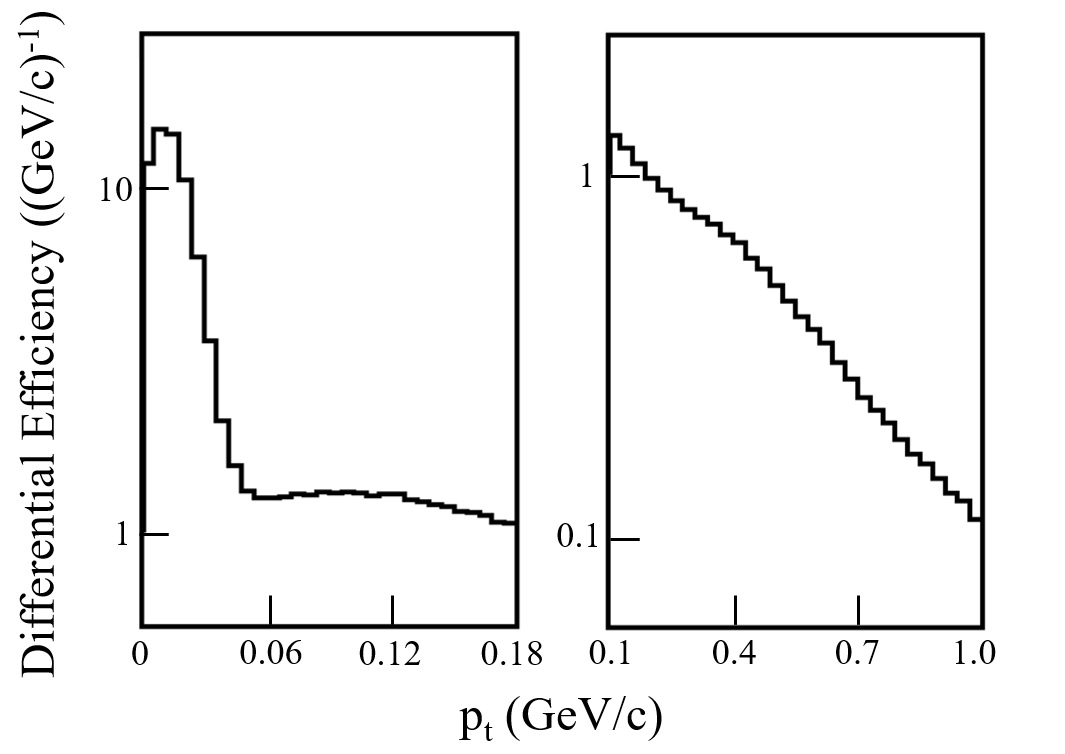}
		\caption{Histograms depicting the number of outgoing protons observed with transverse momentum $p_{t}$. Both were normalized by the total number of particles and bin size to produce the differential efficiency on the y-axis. These histograms are the result of simulating one million protons with initial momentum 3.8 GeV/c and side polarization 1 scattering off of a 51.6 g/cm$^{2}$ block of ethylene, which agree qualitatively with empirical results in Ref.~\cite{azhgirey_measurement_2005}.}
		\label{fig:N-pt}
	\end{figure}

	With the modifications to the code for elastic proton scattering made, it is important to benchmark our code with experimental results to show that everything is working correctly and that the results properly reflect reality. First, the scattering angle part of the differential scattering cross section should be unchanged. 
	To check this, we send in a beam of protons with initial momentum 3.8 GeV/c and side polarization 1 onto a 51.6 g/cm$^{2}$ block of ethylene, a common plastic scintillator. We then record the distribution of the component of momentum transverse to the incident proton direction, $p_{t} = p_{\text{lab}}\sin\theta$, where $p_{\text{lab}}$ is the magnitude of the lab frame proton momentum and $\theta$ the scattering angle. 
	Carrying out a run with one million events, we arrive at the following distribution in Fig.~\ref{fig:N-pt}, which agrees qualitatively with empirical data recorded by Azhgirey et al. in Ref.~\cite{azhgirey_measurement_2005}. We omit a graphical comparison with Ref.~\cite{azhgirey_measurement_2005} due to the inability of Geant4 to reproduce the exact experimental conditions used in the reference.
	
	In Fig.~\ref{fig:N-pt}, it should be noted that the scattering peak at low $p_{t}$ is primarily due to multiple-scattering within the block of ethylene, whereas the gradual drop off afterwards is due to elastic scattering~\cite{azhgirey_measurement_2005}. 
	To analyze only the elastic scattering statistics, we need to enforce a low $p_{t}$ cut in the data. 
	For protons incident with a few GeV of energy, it suffices to cut out data with $p_{t} < 0.1$~GeV/c.
	
\begin{figure}[ht!]
		\centering
		\includegraphics[width=\linewidth]{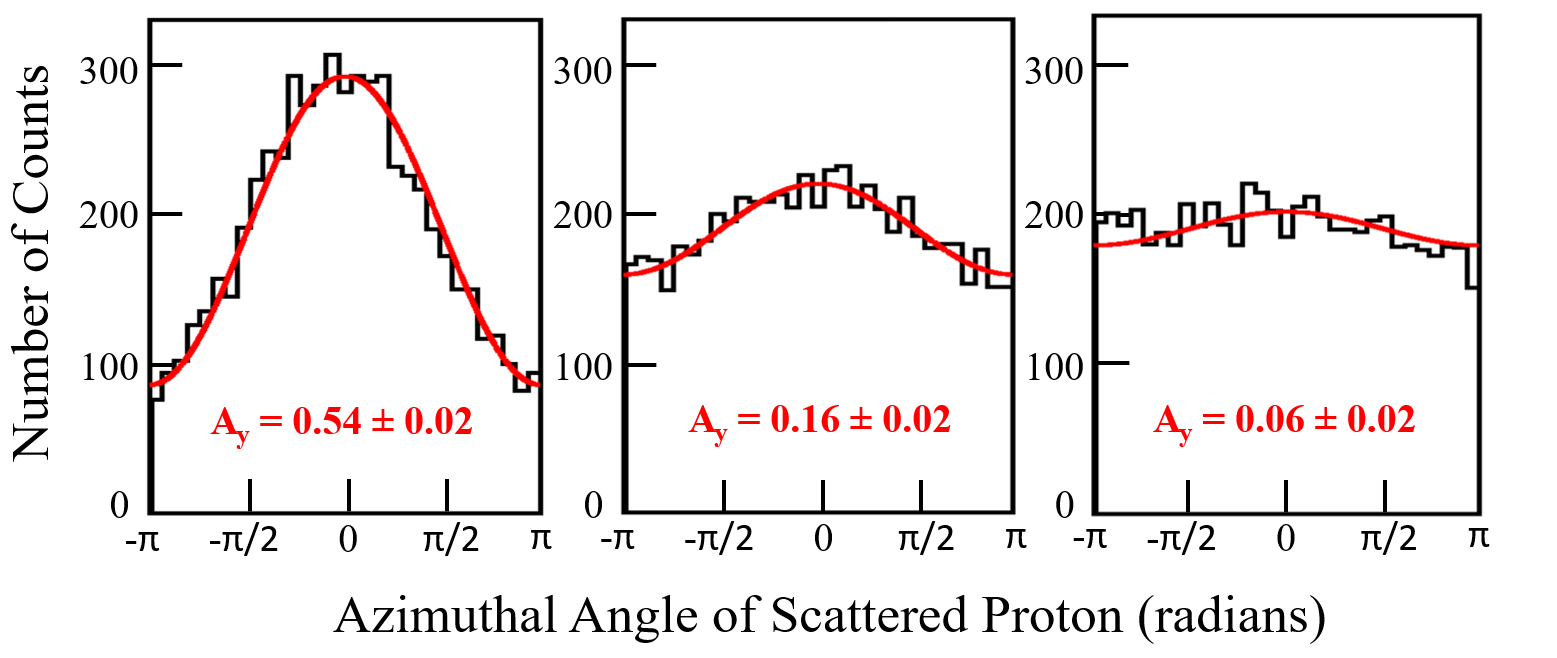}
		\caption{Histograms depicting the number of outgoing protons observed with azimuthal angle $\phi$ overlayed with least squares fits, in red, to determine the analyzing power. For each plot, $10^{5}$ protons were simulated with initial momentum 3.8 GeV/c and side polarization 1 with constant analyzing power set to 0.5, 0.15, and 0.05 respectively from left to right. Color online.}
		\label{fig:N-azim}
	\end{figure}

	To check that our code produces an asymmetry of the correct size in the azimuthal distribution of protons that are elastically scattered off nuclei in the block of ethylene, we test the code with three representative values of the analyzing power. 
	In particular, we generate histograms for the azimuthal angle of scattered protons with $A_{y}$ set to 0.5, 0.15, and 0.05, which are shown in Fig.~\ref{fig:N-azim}. 
	To determine the best fit analyzing power, we perform a least squares fit to the data by choosing $C$ and $A_{y}$ such that
	\begin{equation}
		\sum_{i=1}^{\text{\# of Bins}}\left[N(\phi_{i}) - C(1 + A_{y}P_{y}\cos\phi_{i})\right]^{2} = \text{minimum}.
	\end{equation}
	The resulting fits (shown in Fig.~\ref{fig:N-azim}) yielded analyzing powers of $0.54\pm 0.02$, $0.16\pm 0.02$, and $0.06\pm 0.02$ respectively for the simulations run with $A_{y} =$ 0.5, 0.15, 0.05. Since all of the fits agree within 1-2 standard deviations, we conclude that the azimuthal asymmetry has been properly implemented.
	
	As these checks were made for analyzing powers over the full range of values predicted by the empirical parameterization in Eq.~\ref{eq:Ay}, we proceed with a direct substitution of $A_{y}(\theta)$ where before we had $A_{y}$ equal to a constant.
	
	\section{Results}
	
\begin{figure}[ht!]
		\centering
		\includegraphics[width=\linewidth]{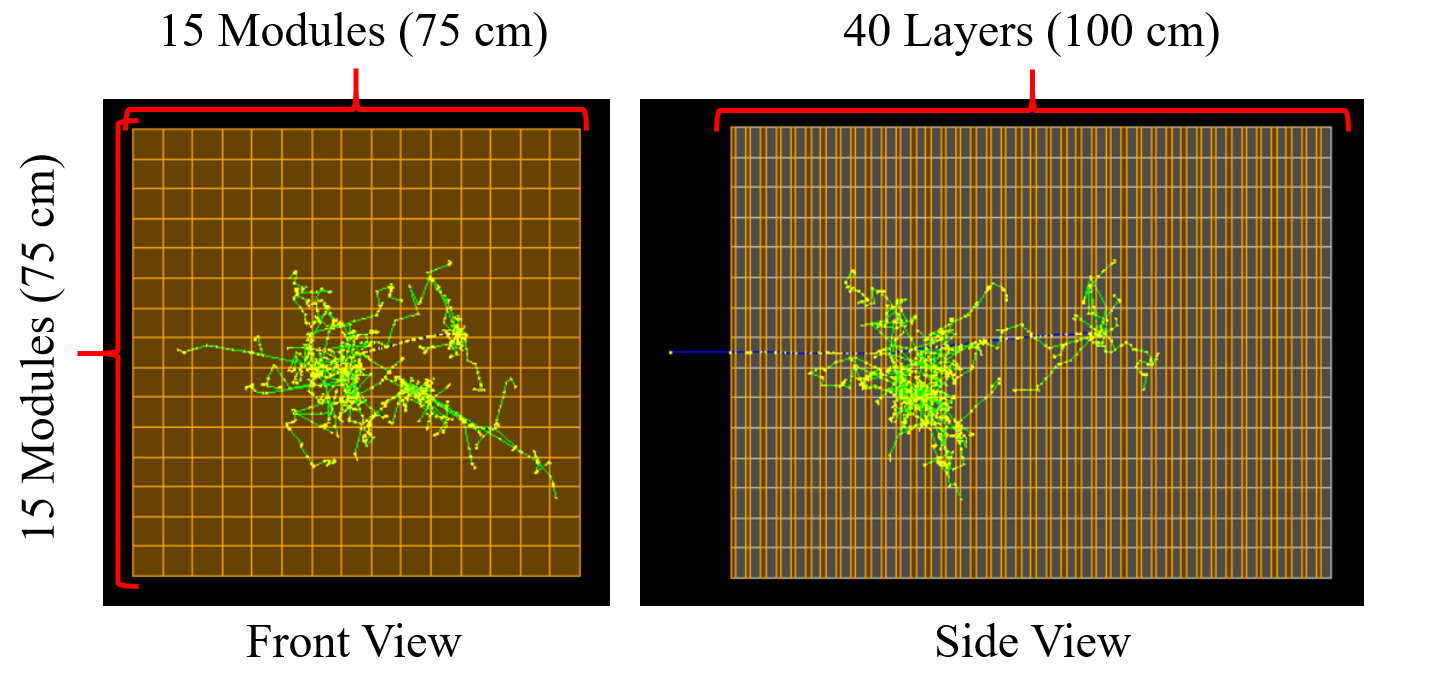}
		\caption{Schematic of the detector used for simulations with a single detection event depicted. 
			The plastic scintillator layers are shaded orange, and iron layers shaded silver. 
			The incident proton is colored in blue, and generated photons are colored in green. Color online.}
		\label{fig:MC}
	\end{figure}

	With the code properly checked and the analyzing power parameterized by Eq.~\ref{eq:Ay}, the shift in detector signal can now be quantified. 
	To mimic as close as possible the calorimeter that would be used for the proposed measurement of the strange form factor of the proton in Ref.~\cite{wojtsekhowski_flavor_2020}, we model the longitudinal structure of our test detector off of the HCAL-J detector at JLab~\cite{franklin_hcalj_2014}, which is similar to the one developed in Ref.~\cite{calo-compass}. 
	In particular, the test detector consists of a 15 by 15 array of $5\times 5\times 100$ cm$^{3}$ modules. Each module is a sampling calorimeter with 40 layers composed of 1.5 cm of Fe and 1 cm plastic scintillator (vinyltoluene). 
	Energy deposits are collected only within the plastic scintillator layers, and position is resolved with a spatial segmentation of 5~cm (see Fig.~\ref{fig:MC}).
	
\begin{figure}[ht!]
		\centering
		\includegraphics[width=0.95\linewidth]{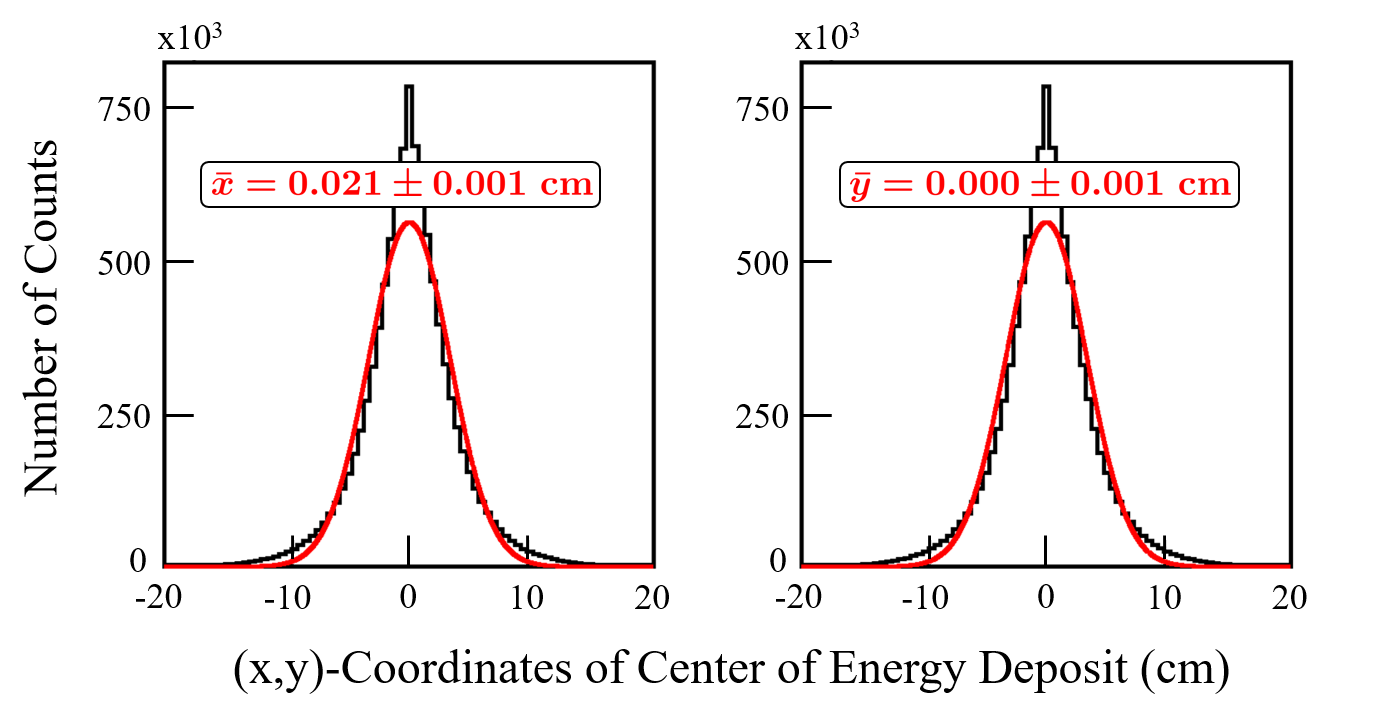}
		\caption{Histograms depicting the center of detector signal for 10 million 2.5 GeV protons with side polarization 1 incident on the described sampling calorimeter. Both distributions were fit to Gaussians shown in red. Color online.}
		\label{fig:shift}
	\end{figure}

	To estimate the shift in detector signal, 10 million 2.5 GeV protons with side polarization 1 were incident on the center of the detector. 
	For each event, the first moment, or center, of energy deposit was computed, with the results accumulated in histograms shown in Fig.~\ref{fig:shift}.
	
	The histograms in Fig.~\ref{fig:shift} were fit to Gaussians, yielding a mean center of detector signal of $\bar{x} = 0.021 \pm 0.001$ cm and $\bar{y} = 0.000 \pm 0.001$ cm. 
	The reported uncertainties only account for statistical uncertainty, neglecting all systematic effects. 
	As expected, the shift is in the direction of the polarization axis crossed with the incident proton direction, or the positive x-axis.
	
	When $A_{y} \, P_{y} << 1$, the shift in detector signal is approximately proportional to the side polarization. Thus, for an arbitrary side polarization $P_{y}$, the expected shift is approximately
	\begin{equation}
		\Delta x \approx P_{y}\times (0.21 \pm 0.01) \text{ mm}.
		\label{eq:deltax}
	\end{equation}

	\section{Discussion and Conclusions}
	
	Ultimately, we found a polarization dependent shift in detector signal on the order of 0.01-0.1 mm given by Eq.~\ref{eq:deltax} for polarized protons incident on a calorimeter. 
	This result is three orders of magnitude smaller than the typical spatial resolution of 2-5~cm in hadronic calorimeters. 
	This small value of the polarization dependent shift is vital for parity violation scattering measurements, as it implies that cuts made in data analysis neglecting this shift in detector signal are still consistent at the level of precision required, for instance, in the proposed experiment in Ref.~\cite{wojtsekhowski_flavor_2020}.
	
	Among other things, these results allow us to begin work on a proposal to the JLab PAC and work on experimental design for a measurement of the strange form factor of the proton at a momentum transfer $Q^{2}$ of 3~(GeV/c)$^{2}$. 
	If shown to be large, we will be provided with yet another clue about the internal dynamics of the quark-gluon sea residing within the proton. 
	Such information is useful for ab initio treatments of the proton and could pave the way for future advances in proton physics.
	
	Additionally, the process by which changes were made to the Geant4 source code in this study are readily generalizable to similar changes that would be needed for empirical corrections to other processes. 
	This would allow for similar spin dependent effects with known azimuthal dependence to be added to Geant4. 
	Such changes are especially useful for the simulation of other parity violating effects, a growing sub-branch of nuclear physics studies~\cite{de_vries_parity-_2020}.
	
	\section{Acknowledgments}
	This work was made possible by the US DOE SULI program at Thomas Jefferson National Accelerator Facility. We would like to extend gratitude towards Dr. John Annand, Dr. Xinzhan Bai, and Dr. Vardan Khachatryan for assistance with Geant4. This work was supported in part by the US DOE Office of Science and Office of Nuclear Physics under the contracts DE-AC02-05CH11231, DE-AC02-06CH11357, and DE-SC0016577, as well as DOE contract DE-AC05-06OR23177, under which JSA, LLC operates JLab.
	
	\bibliography{NIMA}
	
\end{document}